\documentstyle[amssymb,preprint,aps]{revtex}

\newtheorem{theorem}{Theorem}
\newtheorem{lemma}[theorem]{Lemma}
\newtheorem{corollary}[theorem]{Corollary}
\newtheorem{proposition}[theorem]{Proposition}
\newcommand{\proof}{\noindent {\em Proof:}\ \ }
\newcommand{\QED}{\hfill QED.\medskip\par}

\newcommand{\cCo}{{\cal C}_{0}}
\newcommand{\bR}{{\Bbb R}}
\newcommand{\bC}{{\Bbb C}}
\newcommand{\cN}{{\cal N}}
\newcommand{\frt}{{\textstyle{\frac{1}{\sqrt{2}}}}}
\newcommand{\imu}{{\rm i}\,}
\newcommand{\edth}{\eth}

\begin{document}
\draft
\tighten
\preprint{UNE-MSCS-96-128, gr-qc/9705079}
\title{Shear-free Null Quasi-Spherical Spacetimes}
\author{Robert Bartnik 
\thanks{from March 1997: {\tt robertb@ise.canberra.edu.au}}}
\address{
Department of Mathematics, Statistics and Computing Science\\
University of New England\\
Armidale NSW 2351, Australia}
\date{\today}
\maketitle
\begin{abstract}
We study the residual gauge freedom within the null quasi-spherical (NQS) gauge 
for spacetimes admitting an expanding shear-free null foliation.  
By constructing the 
most general NQS coordinates subordinate to such a foliation, we obtain 
both a clear picture of the geometric nature of the residual 
coordinate freedom, and an explicit construction of nontrivial NQS 
metrics representing some well-known spacetimes, such as Schwarzschild, 
accelerated Minkowski, and Robinson-Trautman.  These examples will be 
useful in testing numerical evolution codes.  The geometric gauge freedom 
consists of an arbitrary boost and rotation at each coordinate sphere --- 
and this freedom may be used to normalise the coordinate to an ``inertial''
frame.
\end{abstract}

\section{Introduction}
The recently introduced 
\cite{RAB93,Spillane94,RAB96a,RAB96c}
null quasi-spherical (NQS) coordinate condition 
provides a new approach to the study of the Einstein equations in exterior 
regions admitting an expanding null foliation.
The NQS gauge is described   by the metric ansatz
\begin{equation}
\label{ds2:nqs}
	ds_{NQS}^{2} = -2u\,dz(dr+v\,dz) 
	+ (rd\vartheta+\beta^{1}dr+\gamma^{1}dz)^{2}
	+ (r\sin\vartheta d\varphi + \beta^{2}dr+\gamma^{2}dz)^{2},
\end{equation}
where $(\vartheta,\varphi)$ are the usual polar coordinates on $S^{2}$, 
$u>0$ and $v$ are real-valued functions,  and 
\begin{equation}
	\beta = \beta^{1}\,\partial_{\vartheta} + 
	\beta^{2}\csc\vartheta\,\partial_{\varphi},
\label{def:beta}\quad
	\gamma = \gamma^{1}\,\partial_{\vartheta} + 
	\gamma^{2}\csc\vartheta\,\partial_{\varphi},
\label{def:gamma}
\end{equation}
may be considered either as vectors tangent to the spheres 
$(z,r)=const$ or, using a complex formalism, as spin-1 fields.

The advantages of the gauge, and its generality, are discussed in 
\cite{RAB96c}.  The purpose of this paper is to analyse the 
gauge freedoms remaining within the NQS gauge condition, for the class of 
spacetimes admitting a shear-free ($\beta=0$) null foliation.  
We explicitly describe 
the construction of such foliations in Schwarzschild, Minkowski and 
Robinson-Trautman spacetimes.  

The examples will also be useful as test data for numerical solvers, since they 
involve arbitrary functions but are still simple to describe explicitly.  
This remark applies both to characteristic and 3+1 codes --- the class of
boosted Schwarzschild metrics should be particularly appropriate as test data.

The NQS gauge is best understood by comparison with other popular 
conditions used to describe the metric on a null foliation, due to Bondi
\cite{Bondi62} and Newman and Unti \cite{NU62}. 
 The Newman-Unti radial coordinate $r$ is determined by a 
choice of affine parameter along each of the null generators; the Bondi 
radius is defined by the condition that the spatial volume form 
$\sin\vartheta\,d\vartheta\wedge d\varphi$ have length 
$(r^{2}\sin\vartheta)^{-1}$.
In both cases the angular 
coordinates $(\vartheta,\varphi)$ are transported along the null 
generators.   Both coordinate systems are determined by 
labelling and normalisation conditions at just one transverse $S^{2}$ in 
a null hypersurface and therefore have gauge freedom corresponding to 
functions on a single $S^{2}$ (in each null hypersurface).

By contrast, the NQS radial function $r$ has level sets isometric to  
standard spheres of radius $r$.  Although it is possible to then determine 
the angular coordinates $(\vartheta,\varphi)$  as labelling the outgoing 
null generators (as in \cite{Bondi62},\cite{NU62}), it seems more 
geometrically natural 
to use the $(\vartheta,\varphi)$ determined by the isometry with $S^{2}$ 
with metric $r^{2}(d\vartheta^{2}+\sin^{2}\vartheta\,d\varphi^{2})$. 

Since the metric spheres at each radius are not unique within the null 
hypersurface (at least, this is the case for the standard Minkowski null cone), 
there is
an additional coordinate freedom within the NQS gauge, consisting of a 
choice of Lorentz transformation at each sphere.  This freedom does not
have an analogue in the Bondi and Newman-Unti gauges. 

The vector field $\beta$  is referred to as the {\em 
shear};
that this terminology does not conflict with the accepted usage of ``shear'' 
is seen by noting that the (usual) shear of the null generator $\ell = 
\partial_{r}-r^{-1}\beta$ of the NQS metric Eqn.~(\ref{ds2:nqs}) 
is given in the Newman-Penrose notation by
$\sigma_{NP} = r^{-1}\eth\beta$, where $\eth$ is the eth operator on the 
standard $S^{2}$ and we identify $\beta \sim 
\frac{1}{\sqrt{2}}(\beta^{1}-i\beta^{2})$  with a spin-1 field on $S^{2}$.
Consequently, vanishing shear vector $\beta$ implies vanishing $\sigma_{NP}$;
conversely if $\sigma_{NP}=0$ then $\beta$ consists purely of $\ell=1$ 
spin-1 spherical harmonics \cite{NP66}.  
The role played by the the $\ell=1$ spin-1 
spherical harmonics is discussed in greater detail in the following section.
In the Appendix we show that any shear-free, expanding and twist-free metric admits 
NQS coordinates with $\beta=0$ --- this was shown in \cite{RT62,DIW69} for vacuum metrics.

The metric form (\ref{ds2:nqs}) with $\beta=0$, when
restricted to a coordinate null hypersurface $\cal C$, becomes
\[
	ds^{2}_{\cal C} =	r^{2}(d\vartheta^2+\sin^{2}\vartheta\,d\varphi^{2}).
\]
By identifying ${\cal C}$ with the future null cone at the origin in $\bR^{3,1}$, 
we can see that this form is invariant under the Lorentz group $SO_{0}(3,1)$ --- 
and the Lorentz transformation may also vary with $r$, since invariance 
only requires that each quasi-sphere $r=const.$ is mapped isometrically.
Thus, our main idea is to use explicit representations of the Lorentz group 
acting on the standard null cone ${\cal C}_{0}=\{t=|x|\}$ 
in Minkowski space $\bR^{3,1}$
to describe the general transformation leaving the form $ds^{2}_{\cal C}$
invariant.

Note that the problem of finding general quasi-spherical foliations of a 
null hypersurface which is not shear-free and expanding is considerably 
more difficult, since the explicit model of the standard cone and its 
associated Lorentz deformations is no longer available.  However, 
linearisation arguments suggest strongly that the gauge freedoms of the 
shear-free case are mirrored in the more general setting, provided the 
shear is not too large.  Thus we expect that the description here of the 
shear-free NQS freedom will provide some insight into the more general case.

At least in the shear-free case, we will show that the NQS condition 
has gauge freedom consisting of an $SO_{0}(3,1)$-valued function of the radius 
(on each null hypersurface); this is functionally less rigid than the 
Bondi and NU gauges, since it has  freedom in $r$ which is 
lacking in these gauges.  This Lorentz transformation freedom may be viewed 
as providing a choice of ``inertial frame'' normalisation at each radius, 
and may be used to normalise certain of the remaining metric coefficients, 
as described below.

This interpretation is supported by a comparison \cite{Lun96} between the 
Robinson-Trautman metrics and the NU form \cite{NU63} of the Minkowski metric in 
coordinates using null cones with base point describing a timelike curve.
This comparison may also be made in the NQS coordinates, and supports both 
the  interpretation of Robinson-Trautman spacetimes as describing an 
accelerated black hole rapidly settling down to a Schwarzschild black hole 
in uniform motion, and the interpretation of the NQS freedom as 
representing a choice of reference frame at each radius and time.

In section 2 we study the metric and NQS freedom of the model cone 
${\cal C}_{0}$, by constructing the most general quasi-spherical (QS) 
foliation of ${\cal C}_{0}$.  The resulting metric has shear vector $\beta$ 
consisting solely of $\ell=1$ spherical harmonics, and we show conversely 
that any null surface with such shear vector is gauge-equivalent to the 
standard cone.  In section 3 we describe the metric in  
general NQS coordinates on a spacetime admitting a shear-free null 
foliation.  Section 4 describes the application of these results to the 
specific examples of Schwarzschild, Minkowski and Robinson-Trautman 
spacetimes.  Basic results on shear-free expanding null hypersurfaces are 
collected in the Appendix.

The computations are presented in slightly more detail than is 
strictly necessary, in order to facilitate the use of the example NQS 
metrics in benchmarking numerical codes, and in the interpretation
of general NQS numerical results.

\section{Model cone ${\cal C}_{0}$}
Let ${\cal C}_{0} = \{(x,t)\in \bR^{3,1}: t=|x|\}$ be the standard future null 
cone based at the origin in Minkowski space.
We may use $x\in\bR^{3}$ as a coordinate on ${\cal C}_{0}$; instead a polar 
representation 
\begin{equation}
	x = r\theta,\quad r=|x|, \quad \theta = x/r =(x_{i}/r)  \in S^{2}
\label{rtheta:def}
\end{equation}
will be very useful, where we identify $S^{2}=\{x\in\bR^{3}:|x|=1\}$ and we 
use the direction cosines $\theta=(\theta_{1},\theta_{2},\theta_{3})$, 
$|\theta|=1$ to parameterise $S^{2}$.  The polar coordinates on
$\cCo$ will usually be denoted by $(r,\theta)$ or $(\rho,\zeta)$.

The parameterisation $(\theta_{i})$ of $S^{2}$ leads to a 
representation of tangent vector fields to $S^{2}$ as 3-vector fields on 
the unit sphere
$S^{2}\subset\bR^{3}$ which are tangent to $S^{2}$.  This will prove more 
convenient than using the polar coordinate basis 
$(\partial_{\vartheta},\partial_{\varphi})$.
Thus, a vector field $Y=Y(\theta;\lambda)$, depending on $\theta\in S^{2}$ 
and other parameters $\lambda$ (eg $\lambda=(z,r)$), may be represented as the 
3-vector $Y=(Y_{i})$ satisfying 
$\theta^{T}Y(\theta,\lambda)=\theta_{i}Y_{i}=0$.

Throughout we use latin indices 
$i,j,\ldots$ with range $1,\ldots,3$ and the summation convention on 
repeated indices, not necessarily raised and lowered.

The Minkowski metric induces the rank-2 degenerate 
bilinear form
\begin{equation}
	ds^{2}_{{\cal C}_{0}} = r^{2}|d\theta|^{2} = 
	r^{2}\sum_{i=1}^{3}(d\theta_{i})^{2}
\label{rdth:def}
\end{equation}
on $\cCo$.  Note that $|d\theta|^{2}$ is the standard metric on $S^{2}$, and
$|d\theta|^{2}=\sum_{i,j=1}^{3}\Theta_{ij}dx_{i}dx_{j}$, where $\Theta$ 
is the projection matrix 
\begin{equation}
	\Theta = I - \theta\theta^{T},\qquad 
	\Theta_{ij} = \delta_{ij} - \theta_{i}\theta_{j},
\label{Theta:def}
\end{equation}
and $\theta^{T}$ represents the transpose (row) vector.

A {\em quasi-sphere} of radius $r\in\bR^{+}$ in $\cCo$ is an orientation 
preserving embedding 
$\Phi_{r}:S^{2}\to\cCo$ such that 
\[
	\Phi_{r}^{*}(ds^{2}_{\cCo}) = r^{2}\,|d\theta|^{2},
\]
and we say $\Phi_{r}$ is a {\em quasi-spherical map}.

Every quasi-sphere in $\cCo$ is determined by a unique time and space 
orientation preserving Lorentz transformation
$L\in SO_{0}(3,1)$ via the composition $\Phi_{r} = L\circ i_{r}$
with the inclusion $i_{r}:S^{2}\to\cCo$, 
$\theta\mapsto [r\theta^{T},r]^{T}$,
\begin{equation}
	\Phi_{r}: \theta\in S^{2} \stackrel{i_{r}}{\longmapsto} 
	      \left[ 
	      \begin{array}{c}
	      	r\theta  \\ r
	      \end{array} \right] \in \cCo \stackrel{L}{\longmapsto} 
	L \left[ \begin{array}{c}
	      	r \theta \\ r
	      \end{array} \right] \in \cCo,
\label{Phir:eq}
\end{equation}
and we now exploit this basic description.

The Lorentz transformation $L\in SO_{0}(3,1)$ admits a unique 
boost/rotation decomposition 
$L= {\cal R}{\cal B}$ with
\begin{eqnarray}
\label{rot:def}
	{\cal R} = {\cal R}(R) &=& \left[
	\begin{array}{cc}
		R  & 0  \\ 0&  1
	\end{array}\right], \quad R \in SO(3),
\\
\label{boost:def}
	{\cal B} = {\cal B}(w)&= & \left[
	\begin{array}{cc}
		W & w   \\ w^{T} & b  
	\end{array}\right],\quad w\in\bR^{3},
\end{eqnarray}
where $b=\sqrt{1+|w|^{2}}\ge1$ and $W$ is the $3\times3$ 
matrix
\begin{equation}
	W = I + \frac{1}{b+1} ww^{T}.
\label{W:def}
\end{equation}
It is easily checked that $\cal B$ preserves the Minkowski metric $\eta$,
$\eta = {\cal B}^{T}\eta {\cal B}$.  The quasi-spherical  map 
$\Phi_{r}$ may be described explicitly by
\begin{equation}
	\Phi_{r}(\theta) = L \left[ 
	      \begin{array}{c}
	      	r\theta  \\ r
	      \end{array} \right] = \left[ 
	      \begin{array}{c}
	      	rR\theta+r\left(1+\frac{\theta^{T}w}{b+1}\right) Rw \\
	      	r(b+\theta^{T}w)  
	      \end{array} \right],
\label{Phi1:eq}
\end{equation}
where $\theta^{T}w= w^{T}\theta $ is the usual inner product between 
column 3-vectors.
In terms of the rectangular $\bR^{3,1}$ parameterisation of  the target $\cCo$ we 
have $\Phi_{r}(\theta) = [\rho(r,\theta)\zeta(\theta)^{T},\rho(r,\theta)]^{T}$, 
where we define
\begin{eqnarray}
\label{f:def}
	f(\theta)= f(\theta;w)  &:=&   b+\theta^{T}w = \sqrt{1+|w|^{2}} + \theta^{T}w,
\\
\label{rho:def}
	\rho(r,\theta) = \rho(r,\theta;w)& := & r f(\theta) 
\\
\label{zeta:def}
	\zeta(\theta)  =  \zeta(\theta;w,R) & := &  f^{-1}R
	  \left(\theta + \left(1+\textstyle{\frac{\theta^{T}w}{b+1}}\right)w\right) 
	  = f^{-1} R(w+W\theta).
\end{eqnarray}
Here and elsewhere we adopt the convention that $(\rho,\zeta)$ denotes 
polar coordinates on the range (target) $\cCo$ of $\Phi_{r}$, which leads 
to the description $\Phi_{r}(\theta)=(\rho(r,\theta),\zeta(\theta))$ of 
$\Phi_{r}$ in polar coordinates.
The metric $ds^{2}_{\cCo} $ on the target cone $\cCo$ in polar coordinates 
$(\rho,\zeta)$ is just $\rho^{2}|d\zeta|^{2}$ and we may verify the 
quasi-spherical condition $\Phi_{r}^{*}(ds^{2}_{\cCo}) = 
r^{2}|d\theta|^{2}$ by direct computation as follows.  

We define the angular gradient operator $D_{\theta}=(D_{\theta_{i}})$ as 
the projection tangent to the unit sphere
 $S^{2}$ of the ordinary gradient in $\bR^{3}$.  
Explicitly, let $h_{e}(x) := h(x/|x|)$, $x\in\bR^{3}$ be the 
homogeneous degree 0 extension of any $h\in C^{1}(S^{2})$ and define
$D_{\theta_{i}}h = \partial h_{e}/\partial x_{i}$.  It follows that 
$\theta_{i}D_{\theta_{i}}h= 0$ by the homogeneity condition, 
and $D_{\theta_{i}}\theta_{j}=\Theta_{ij}$.
Then $\Phi_{r}^{*}(\rho^{2}|d\zeta|^{2}) = 
\rho(r,\theta)^{2}|\zeta_{,\theta}d\theta|^{2}$,
where $(\zeta_{,\theta})_{ij} = D_{\theta_{j}}\zeta_{i}$ with
\begin{eqnarray}
	D_{\theta_{j}}\zeta_{i}
	&=&  D_{\theta_{j}} \left(%
	(b+\theta^{T}w)^{-1 }R_{ik}\left(w_{k} + W_{kl}\theta_{l}\right)\right)
\nonumber\\
	 & = & f^{-1} R_{ik}\left(-f^{-1}(w_{k}+W_{kl}\theta_{l})\Theta_{jm}w_{m}+ 
	      W_{km}\Theta_{mj}\right)
\nonumber\\
\label{dzetadtheta:eq}
	 & = & f^{-1}R_{ik}A_{km}\Theta_{mj},
\end{eqnarray}
and we have introduced the very useful matrix 
\begin{equation}
	A := I - f^{-1}\left( \theta + (b+1)^{-1}w\right) w^{T}.
\label{A:def}
\end{equation}

By exercising a certain amount of care we may convert to a matrix notation.
We adopt the 
convention that 3-vector quantities such as $\theta_{i}$, $d\theta_{i}$, $w_{i}$, 
are to be treated as 
column vectors (of 1-forms, functions, etc.) and 
that row vectors usually will be  indicated by the transpose 
notation, {\em except} that we regard $D_{\theta}$, 
$D_\zeta$ and their associated gradients as {\em row} 
vectors.

For example, with these conventions the computation 
(\ref{dzetadtheta:eq}) may be summarised as
\begin{equation}
	\zeta_{,\theta} = f^{-1}RA \Theta  
\label{dzeta:eq}
\end{equation}
and the chain rule for $h\in C^{\infty}(S^{2},\bR)$ appears as
\begin{eqnarray*}
	dh & = & h_{,\theta}\, d\theta
\\
    d (h\circ\zeta) &=& h_{,\zeta}\zeta_{,\theta}\,d\theta
\\
	 (h\circ\zeta)_{,\theta} & = & 
	f^{-1} h_{,\zeta} RA\Theta,
\end{eqnarray*}
where both sides of the final identity are row vectors.

The matrix $A$ satisfies several useful identities,
\begin{eqnarray}
\label{ATA:eq1}
	A^{T}A & = & I - f^{-1}(w\theta^{T}+ \theta w^{T}),
\\
\label{ATA:eq2}
	\Theta A^{T}A \Theta & = & \Theta,
\\
\label{Ainv:eq}
	A^{-1} & = & I + (\theta + (b+1)^{-1}w)w^{T} = W+\theta w^{T},
\end{eqnarray}
which may be used to show that $\Phi_{r}$ is quasi-spherical:
\begin{eqnarray*}
	\Phi_{r}^{*}(\rho^{2}|d\zeta|^{2}) & = & r^{2}f^{2} |f^{-1}RAd\theta|^{2}
	\ =\ r^{2}\,d\theta^{T} A^{T} R^{T} R A \,d\theta
\\
	 & = & r^{2}\,d\theta^{T} A^{T}A\,d\theta = r^{2}|d\theta|^{2},
\end{eqnarray*}
since $R^{T}R=I$ and $\theta^{T}d\theta = 0$, so $d\theta=\Theta d\theta$.
This confirms that $\Phi_{r}$ is quasi-spherical and also conformal, since
$\Phi^{*}_{r}(|d\zeta|^{2}) = f^{-2}|d\theta|^{2}$.
Note that from Eqs.~(\ref{zeta:def}),(\ref{Ainv:eq}) we have 
\begin{equation}
	\zeta(\theta) = f^{-1} R A^{-1T}\theta,
\label{zeta:eq}
\end{equation}
where $A^{-1T} $ is the inverse transpose matrix.

Having described a single quasi-sphere, we may now consider the effects of 
$r$-dependence:  a map $\Phi:\cCo \to \cCo$ is said to be {\em 
quasi-spherical} if $\Phi_{r} = \Phi\circ i_{r}$ is a quasi-spherical map 
for each $r>0$, and if $r \mapsto \Phi_{r}$ is at least continuously 
differentiable in $r$ --- although for simplicity we shall consider only 
smooth maps.  Equivalently, $\Phi(r,\theta) = L(r)\circ 
i_{r}(\theta) $, where the Lorentz transformations 
$L(r) = {\cal R}(R(r)) {\cal B}(w(r)) $ are described by boost 
and rotation maps $w\in C^{\infty}(\bR^{+},\bR^{3})$, $R\in C^{\infty}(\bR^{+},SO(3))$.
Note that we do not require that $\Phi$ be a diffeomorphism, and this does 
not follow from the condition $	\Phi^{*}(ds^{2}_{\cCo}) = r^{2}\,|d\theta|^{2}$
since $ds^{2}_{\cCo}$ is degenerate.  
Note also that $\Phi^{-1}$, when defined, is not usually quasi-spherical; 
neither is the composition of two quasi-spherical maps usually again 
quasi-spherical.

The general quasi-spherical map $\Phi:\cCo \to \cCo$ is thus described by
\[
	\Phi(r,\theta) = {\cal R}(R(r)) {\cal B}(w(r)) 
	\left[ \begin{array}{c} r \theta \\ r	\end{array} \right]
	= \left[ \begin{array}{c} 
	    \rho(r,\theta)\zeta(r,\theta)  \\ \rho(r,\theta)	
	         \end{array} 
	  \right]
\]
where the functions $\rho(r,\theta),\zeta(r,\theta)$ are defined by 
Eqs.~(\ref{rho:def}),(\ref{zeta:def}) with $w,R$ now depending on $r$.  
The pull-back metric is then
\begin{eqnarray*}
	\Phi^{*}(ds^{2}_{\cCo}) & = & 
	\rho(r,\theta)^{2}|\zeta_{,\theta}d\theta+\zeta_{,r}dr|^{2}
\\
	 & = & \rho^{2}|\zeta_{,\theta}d\theta|^{2} + 2\rho^{2} 
	 d\theta^{T}\zeta_{,\theta}^{T}\zeta_{,r}dr + \rho^{2}|\zeta_{,r}|^{2}dr^{2},
\end{eqnarray*}
where $\zeta_{,r}$ denotes the column vector of partial derivatives
$\frac{\partial}{\partial r}\zeta$.
Note that we are taking the liberty of using $\rho,\zeta$  to denote both the 
coordinates $\rho,\zeta$ on $\cCo$ and their pullbacks $\rho(r,\theta) = 
\Phi^{*}(\rho)$, $\zeta(r,\theta) = \Phi^{*}(\zeta)$, which are  
functions determining the map $\Phi$ 
--- this ambiguity should not lead to any serious confusion.

We have already checked that 
$\rho^{2}|\zeta_{,\theta}d\theta|^{2}=r^{2}|d\theta|^{2} $, 
and we compute from Eqn.~(\ref{zeta:eq})
\begin{equation}
	fA^{-1}R^{-1}\zeta_{,r} = A^{-1}R^{-1}R_{,r}A^{-1T}\theta
	+ A^{-1} \frac{\partial}{\partial r}(A^{-1T})\theta 
	- f^{-1} f_{,r} (A^{T}A)^{-1}\theta.
\label{zetadr:eq}
\end{equation}
The first term on the right hand side is simplified by introducing the 
antisymmetric matrix 
\begin{equation}
	S_{1} := WR^{-1}R_{,r}W,
\label{S1:def}
\end{equation}
and equals (bearing in mind the formulas
$A^{-1}W^{-1} = I+b^{-1}\theta w^{T}$, $W^{-1}=I - b^{-1}(b+1)^{-1}ww^{T}$)
\[
	S_{1}\theta + b^{-1}\Theta S_{1}w.
\]
After some computation, we find that 
the second and third terms of Eqn.~(\ref{zetadr:eq}) may be combined into
\[
	\Theta W^{-1}w_{,r} + \frac{1}{b+1} \theta\times(w_{,r}\times w),
\]
where $a\times b = (\epsilon_{ijk}a_{j}b_{k})$ is the usual cross product 
in $\bR^{3}$.
Defining the 3-vectors $s_{1},t_{1}$ (depending on $r$ but independent of $\theta$)
\begin{eqnarray}
\label{s1:def}
	s_{1} & := & *S_{1} + \textstyle{\frac{1}{b+1}}w_{,r}\times w
\\
\label{t1:def}
	t_{1} & := & b^{-1}S_{1}w + W^{-1}w_{,r}
\end{eqnarray}
where $*S := (\frac{1}{2}\epsilon_{ijk}S_{jk})$, and the 
$(r,\theta)$-dependent vector $\beta$
\begin{equation}
	\beta := r\Theta t_{1} + r \theta \times s_{1},
\label{beta:def}
\end{equation}
 we have the identity
\begin{equation}
	\rho \zeta_{,r} = R A \beta.
\label{zetadr:id}
\end{equation}
This may be used to simplify $\Phi^{*}(ds^{2}_{\cCo})$:
\begin{eqnarray*}
	\rho^{2}d\theta^{T}\zeta_{,\theta}^{T}\zeta_{,r}dr & = & 
	r d\theta^{T}\,A^{T}R^{T}RA\beta\,dr
\\
	 & = & r d\theta^{T} A^{T}A \beta\, dr = r d\theta^{T}\beta\,dr,
\end{eqnarray*}
since $\theta^{T}\beta = 0$.  Similarly we find 
\[
	\rho^{2}|\zeta_{,r}|^{2} = \beta^{T}A^{T}R^{T}RA\beta = |\beta|^{2},
\]
and it follows that $\Phi^{*}(ds^{2}_{\cCo})= |r d\theta + \beta dr|^{2}$,
which is in quasi-spherical form with shear vector $\beta$.

To summarise:

\begin{proposition}
Suppose $\Phi:\cCo\to\cCo$ is a $C^{\infty}$ quasi-spherical map.  Then $\Phi$ 
satisfies 
Eqn.~(\ref{Phi1:eq}) for some  Lorentz boost parameter $w\in 
C^{\infty}(\bR^{+},\bR^{3})$ and spatial rotation $R\in C^{\infty}(\bR^{+},SO(3))$.
In the rectangular-polar coordinates $(r,\theta)$ on $\cCo$ we have
\begin{equation}
	\Phi^{*}(ds^{2}_{\cCo}) = |r d\theta + \beta dr|^{2},
\label{cone:nqs}
\end{equation}
where the shear vector $\beta$ is defined in terms of $w$ and $R$ by 
Eqs.~(\ref{S1:def}),(\ref{s1:def}),(\ref{t1:def}),(\ref{beta:def}).
\end{proposition}

The angular vector field $\beta$ consists solely of spin-1 $\ell=1$ spherical harmonics.
This follows from Eqn.~(\ref{beta:def}), since for any constant vector 
$t\in\bR^{3}$, the angular vector fields $\Theta t = t -(\theta^{T}t)\,\theta$
and $\theta\times t$ satisfy 
\begin{eqnarray}
\label{grad:eq}
	\Theta t & = & \mbox{grad}_{S^{2}}(\theta^{T}t),	
\\
\label{Jgrad:eq}
	\theta\times t & = & J \mbox{grad}_{S^{2}}(\theta^{T}t),
\end{eqnarray}
where the complex structure $J:TS^{2}\to TS^{2}$ is defined by 
anticlockwise rotation with respect to the outer normal $\theta$ to 
$S^{2}\subset\bR^{3}$.  
Now if $e_{1},e_{2}$ is an 
oriented orthonormal frame on $S^{2}$ (so $e_{1}\times e_{2}=\theta$), 
we define the spin-1 projection of 
a vector field $X=X^{1}e_{1}+X^{2}e_{2}$ by
\begin{equation}
	X\ \sim\ \xi = \frt(X^{1}-\imu X^{2})
\label{spin1:def}
\end{equation}
(note $JX=\theta\times X=-X^{2}e_{1}+X^{1}e_{2}\sim -\imu \xi$), 
and the operator eth by 
\begin{equation}
	\eth = \frt(\nabla_{e_{1}} - \imu \nabla_{e_{2}}),
\label{eth:def}
\end{equation}
where $\nabla$ is the standard covariant derivative on $S^{2}$.
Note that as so defined, $\xi$ and $\edth$ are frame dependent, hence the 
use of $\sim$.  Defining the basis vector $e=\frt(e_{1}-\imu e_{2})$, we 
could instead write the equality $X=\bar{\xi}e+\xi\bar{e}$ and then 
consider $\xi$ as the coefficient of the representation of $X$ as a section 
of a spin-1 complex line bundle, with respect to the basis vector $e$.

If $\phi,\psi\in C^{1}(S^{2},\bR)$ then 
\[
	\eth(\phi+\imu\psi) = \frt (\phi_{,1} + \psi_{,2} 
	+\imu(\psi_{,1}-\phi_{,2})),
\]
where the subscripts $(\cdot)_{,a}$ for  $a=1,2$ denote directional derivatives with respect to 
the basis vectors $e_{1},e_{2}$,
and the vector field correspondence is
\[
	\mbox{grad}_{S^{2}}\phi - J\mbox{grad}_{S^{2}}\psi 
   \  \sim\ \eth(\phi+\imu\psi)  .
\]
In particular, for any constant $s,t\in\bR^{3}$ we have the correspondence
\begin{equation}
	\Theta t + \theta\times s\ \sim\ \eth(\theta^{T}(t-\imu s)),
\label{ethl=1}
\end{equation}
and the identity 
\begin{equation}
	\Delta_{S^{2}}(\theta^{T}t) = -2\, \theta^{T}t
\label{Lap:l=1}
\end{equation} 
completes the identification of  $\Theta t + \theta\times s$ as a spin-1 $\ell=1$
spherical harmonic.

We also derive from Eqn.~(\ref{Lap:l=1}) that
\begin{equation}
	\Delta_{S^{2}}((\theta^{T}t)^{2}-{\textstyle\frac{1}{3}}|t|^{2}) 
	= -6 (\,(\theta^{T}t)^{2}-{\textstyle\frac{1}{3}}|t|^{2})
\label{Lap:l=2}
\end{equation}
and thus $\theta^{T}t\,\theta^{T}s - {\textstyle\frac{1}{3}}t^{T}s$ is an 
$\ell=2$ spherical harmonic for any $s,t\in\bR^{3}$.

The relations (\ref{s1:def}),(\ref{t1:def}) between $s_{1},t_{1}$ and 
$S_{1},w_{,r}$ may be inverted, since by direct computation we have the 
following lemma.

\begin{lemma} \label{lemma2}
Suppose $s,t,w,\tilde{w},\sigma\in\bR^{3}$ and let $b:=\sqrt{1+|w|^{2}}$ 
and $W^{-1} = I-\frac{1}{b(b+1)}ww^{T}$.  The equations
\begin{equation}
\left\{	\begin{array}{rcl}
		t &=& b^{-1}w\times\sigma + W^{-1}\tilde{w}  \\
		s &=& \sigma + \frac{1}{b+1}\tilde{w}\times w
	\end{array}\right.
\label{st:lem1}
\end{equation}
are equivalent to 
\begin{equation}
\left\{	\begin{array}{rcl}
		\tilde{w} & = & b t + s\times w  \\
		\sigma  & = & bW^{-1}s + \frac{b}{b+1} w\times t.
	\end{array}\right.
\label{st:lem2}
\end{equation}
\end{lemma}

Applying this lemma with $s=s_{1}$, $t=t_{1}$, $\tilde{w}=w_{,r}$, 
$\sigma=*S_{1}$ and assuming (\ref{s1:def}),(\ref{t1:def})
gives the ordinary differential equations
\begin{eqnarray}
\label{wr:ode}
	w_{,r} &=& b t_{1} + s_{1}\times w
\\
\label{Rr:ode}
	R_{,r} & = & R W^{-1} (b\,{*}(W^{-1}s_{1}) + \textstyle{\frac{b}{b+1}}
	\,{*}(w\times t_{1})) W^{-1},
\end{eqnarray}
where for any vector $t$ we define the matrix $*t = (\epsilon_{ijk}t_{k})$.
(So the two star operations interchange vectors with antisymmetric 
matrices.)
Consequently we have the following reconstruction results.

\begin{proposition} \label{wRode:prop}
Suppose $s_{1},t_{1}\in C^{\infty}(\bR^{+},\bR^{3})$ are given functions.  
Given initial conditions 
\begin{equation}
	\begin{array}{rcl}
		w(r_{0}) & = & w_{0}\ \in\bR^{3}   \\
		R(r_{0}) & = & R_{0}\ \in SO(3),
	\end{array}
\label{wR0:ivp}
\end{equation}
there is a 
unique quasi-spherical map $\Phi:\cCo\to\cCo$ with parameters $w\in 
C^{\infty}(\bR^{+},\bR^{3})$, $R\in C^{\infty}(\bR^{+},SO(3))$ satisfying 
$w(r_{0})=w_{0}$, $R(r_{0})=R_{0}$ and such that the shear vector $\beta$
satisfies 
\[
	\beta(r,\theta) = \Theta t_{1}(r) + \theta\times s_{1}(r).
\]
\end{proposition}

\proof
With $s_{1}, t_{1}$ given functions of $r$ and $b=\sqrt{1+|w|^{2}}$, 
Eqn.~(\ref{wr:ode}) gives an ordinary differential equation for $w$,
with initial condition $w(r_{0})=w_{0}$.
Since for $r\in[r_{0},r_{1}]$
\[
	 |b(r) t_{1}(r) + s_{1}(r)\times w(r)| 
	\le 2  \,(|w(r)| + 1)\sup_{[r_{0},r_{1}]}(|t_{1}| + |s_{1}|) ,
\]
by Gronwall's inequality the solution of Eqn.~(\ref{wr:ode}) is locally bounded and may be continued to all 
$r\in\bR^{+}$.  Substituting $w(r)$ into Eqn.~(\ref{Rr:ode}) gives an 
ode for $R(r)\in SO(3)$ with initial condition $R(r_{0})=R_{0}$, 
which similarly has a global solution $R\in C^{\infty}(\bR^{+},SO(3))$.  
The quasi-spherical map defined by the solutions $w(r), R(r)$  
via Eqn.~(\ref{Phi1:eq}) has the required shear vector $\beta$ by 
previous computations.  $\Phi$ is the unique map satisfying the initial 
conditions at $r_{0}$ since any quasi-spherical map may be put into the 
form (\ref{Phi1:eq}) and the parameters are then uniquely determined by 
the initial value problem (\ref{wr:ode}),(\ref{Rr:ode}),(\ref{wR0:ivp}).
\QED

\begin{corollary}
Suppose $|r d\theta + \beta dr|^{2}$ is a quasi-spherical form on $\cCo$, 
with shear vector $\beta$ consisting solely of $\ell=1$ spherical harmonics
(ie.~$\beta$ may be expressed in the form  Eqn.~(\ref{beta:def})).
Then there is a quasi-spherical map $\Phi:\cCo\to\cCo$, 
$\Phi(r,\theta)=(\rho,\zeta)$, such that 
\begin{equation}
	\Phi^{*}(\rho^{2}|d\zeta|^{2}) = |r d\theta + \beta dr|^{2}. 
\label{ell1=gauge}
\end{equation}
Furthermore, $\Phi$ is unique up to a rigid Lorentz transformation of 
$\cCo$: if $\tilde{\Phi}:\cCo\to\cCo$ is any map satisfying 
Eqn.~(\ref{ell1=gauge}), then there is 
$L_{0}\in SO_{0}(3,1)$ such that $\tilde{\Phi}=L_{0}\circ\Phi$.
\end{corollary}

\proof
Since $\beta$ is pure $\ell=1$, the $\ell=1$ spherical harmonic coefficient
functions $s_{1}(r), t_{1}(r)$ are uniquely determined by 
Eqn.~(\ref{beta:def}), and an appropriate quasi-spherical map $\Phi$ may be constructed 
using Proposition \ref{wRode:prop} and initial conditions $w(r_{0})=0$,
$R(r_{0})=I$ at some radius $r_{0}$.
If $\tilde{\Phi}:\cCo\to\cCo$ also satisfies Eqn.~(\ref{ell1=gauge})
then $\tilde{\Phi}$ is quasi-spherical and hence may be parameterised by 
Lorentz transformations $\tilde{L}(r)$, with parameter functions 
$\tilde{w}(r)$ and $\tilde{R}(r)$.  Let $w_{0}=\tilde{w}(r_{0})$ and 
$R_{0}=\tilde{R}(r_{0})$ and let $L_{0}$ be the corresponding Lorentz 
transformation.  Because $L_{0}^{*}(r^{2}|d\theta|^{2}) = 
r^{2}|d\theta|^{2}$, the map $L_{0}\circ\Phi$ is also quasi-spherical 
satisfying
Eqn.~(\ref{ell1=gauge}), and 
has parameters $\hat{w}(r)$, $\hat{R}(r)$ 
satisfying the initial conditions $\hat{w}(r_{0}) = w_{0}$, 
$\hat{R}(r_{0})=R_{0}$.  
Since the parameters $s_{1}, t_{1}$ are determined uniquely from $\beta$ 
in Eqn.~(\ref{ell1=gauge}),  uniqueness of the solution of the initial 
value problem Eqs.~(\ref{wr:ode}),(\ref{Rr:ode}),(\ref{wR0:ivp}) implies 
$\tilde{\Phi}=L_{0}\circ\Phi$ as required.  
\QED

Note that $\tilde{w}(r)$, 
$\tilde{R}(r)$ can be computed in terms of $w(r)$, $R(r)$ and $w_{0}, 
R_{0}$ from the identity
\[
	\tilde{L}={\cal R}(\tilde{R}){\cal B}(\tilde{w})
	= {\cal R}(R_{0}){\cal B}(w_{0}){\cal R}(R){\cal B}(w).
\]
However it is not true in general that the composition of quasi-spherical 
maps is again quasi-spherical --- $\tilde{L}$ in this identity defines a 
quasi-spherical map only when $w_{0}$, $R_{0}$ are constant.

\section{Deformation of spacetime metrics}

We consider now those spacetimes whose metric can be placed in the form
\begin{equation}
	ds^{2}_{SF} =  -2U\,dz(dr + V\,dz) + |r\,d\theta + \Gamma\,dz|^{2}, 
\label{ds2SF:def}
\end{equation}
where $\Gamma=\Gamma(z,r,\theta)$ is an angular vector field, 
so $\Gamma$ satisfies $\theta^{T}\Gamma= 0 $.  We may verify that the 
null congruence defined by the coordinate tangent vector $\partial_{r}$ is
expanding, shear-free and twist-free.
This class includes the Schwarzschild, Robinson-Trautman and accelerated 
Minkowski spacetimes, and will be further discussed in the following section.
For the present, we use the techniques of the previous section 
to study the effect of 
quasi-spherical Lorentz deformations of metrics of the form 
(\ref{ds2SF:def}).

We regard $ds^{2}_{SF}$ as defined on (a subset of) $\bR\times\cCo$, and 
then the metric induces the standard form $r^{2}|d\theta|^{2}$ on $\cCo$.
The form (\ref{ds2SF:def}) is in fact the most general metric form 
compatible with this property and such that the coordinate $z$ is null 
(characteristic).
Extending previous definitions, for any domain $\Omega\subset\bR\times\bR^{+}$
with coordinates $(z,r)$, we say that $\Phi:\Omega\times S^{2}\to 
\bR\times\cCo$ is {\em quasi-spherical} if  the 
restrictions $\Phi_{(z,r)}$ map $S^{2}\to \cCo$ and are quasi-spherical, for 
each $(z,r)\in\Omega$.
As previously, we shall assume $\Phi$ is $C^{\infty}$.

In order to compute the pullback $\Phi^{*}(ds^{2}_{SF})$ using the above 
techniques, we first rename the polar coordinates in 
Eqn.~(\ref{ds2SF:def}) from $(r,\theta)$ to $(\rho,\zeta)$.  Thus we now 
regard the metric parameters $U,V,\Gamma$ as functions of the coordinates 
$(z,\rho,\zeta)$ on the range $\bR\times\cCo$ of $\Phi$, and
we reserve $(z,r,\theta)$ for coordinates on the domain $\Omega\times S^{2}$. 

The map $\Phi$ may be described using the Lorentz boost and rotation 
functions $w\in C^{\infty}(\Omega,\bR^{3})$, $R\in C^{\infty}(\Omega,SO(3))$ via
\begin{equation}
	\Phi(z,r,\theta) = (z,\rho,\zeta) = (z,\rho(z,r,\theta),\zeta(z,r,\theta)),
\label{Phi:def2}
\end{equation}
where as before,
\begin{eqnarray}
\label{rho:def2}
	\rho(z,r,\theta) & = & r\,f(\theta;w(z,r)) = r\,(\sqrt{1+|w|^{2}} + 
	\theta^{T}w),
\\
\label{zeta:def2}
	\zeta(z,r,\theta) & = & \zeta(\theta;w(z,r),R(z,r)) = f^{-1}R (w+W\theta).
\end{eqnarray}
To compute the pullback $\Phi^{*}(ds^{2}_{SF})$ we use Eqn.~(\ref{wr:ode})
and the definitions 
Eqs.~(\ref{S1:def}), (\ref{s1:def}), (\ref{t1:def}), (\ref{beta:def}):
\begin{eqnarray}
	\rho_{,r} & = & \frac{\partial \rho}{\partial r}
	            = f + r (\theta+b^{-1}w)^{T}w_{,r}
\nonumber\\
	 & = & f + r (\theta+b^{-1}w)^{T}(bt_{1}+ s_{1}\times w)
\nonumber\\
	 & = & f + r w^{T}(\Theta t_{1}+ \theta\times s_{1}) 
	         + r (b+w^{T}\theta)\theta^{T}t_{1}
\nonumber\\
     & = & f (1+r\theta^{T}t_{1}) + w^{T}\beta.
\label{drhodr:eq}
\end{eqnarray}
Defining the 3-vector quantities $s_{0},t_{0},\hat{\gamma}$ by
\begin{eqnarray}
\label{S0:def}
	S_{0} & := & WR^{-1}R_{,z}W
\\
\label{s0:def}
	s_{0} & := & *S_{0} + {\textstyle \frac{1}{b+1}} w_{,z}\times w
\\
\label{t0:def}
	t_{0} & := & b^{-1}S_{0}w + W^{-1}w_{,z} = b^{-1}(w_{,z}-s_{0}\times w)
\\
\label{gamhat:def}
	\hat{\gamma} & := & r(\Theta t_{0} + \theta\times s_{0}),
\end{eqnarray}
we similarly find
\begin{equation}
	\rho_{,z} = r f_{,z} = rf\theta^{T}t_{0} + w^{T}\hat{\gamma}
\label{drhodz:id}
\end{equation}
and thus
\begin{equation}
	\Phi^{*}(d\rho) = (f(1+r\theta^{T}t_{1})+w^{T}\beta)\,dr 
	        + (rf\theta^{T}t_{0}+w^{T}\hat{\gamma})\,dz 
	        + rw^{T}d\theta.
\label{Phi*drho:eqn}
\end{equation}
Using the identities Eqs.~(\ref{dzeta:eq}), (\ref{zetadr:id}) and
the analogous
\begin{equation}
	\rho \zeta_{,z}= RA\hat{\gamma}, 
\label{dzetadz:eq}
\end{equation}
we also have
\begin{equation}
	\Phi^{*}(d\zeta) = \zeta_{,r}dr+\zeta_{,z}dz+\zeta_{,\theta}d\theta
	 = \rho^{-1}RA(\beta dr + \hat{\gamma}dz + r\,d\theta).
\label{Phi*dzeta:eq}
\end{equation}

We denote the pullbacks of the metric functions $U,V,\Gamma$ by a tilde, 
so for example $\tilde{U}=\Phi^{*}(U)$ and $\tilde{U}(z,r,\theta) = 
U(z,\rho(z,r,\theta),\zeta(z,r,\theta))$.
Substituting Eqs.~(\ref{Phi*drho:eqn},\ref{Phi*dzeta:eq}) into 
Eqn.~(\ref{ds2SF:def}) with coordinates $(\rho,\zeta)$ replacing $(r,\theta)$ 
as already mentioned, gives
\begin{eqnarray}
	\Phi^{*}(ds^{2}_{SF}) & = & 
	{}-2\tilde{U} (f(1+r\theta^{T}t_{1})+w^{T}\beta)\,dz\,dr
\nonumber\\
	&&{}-2\tilde{U} (\tilde{V}+rf\theta^{T}t_{0}+w^{T}\hat{\gamma})\,dz^{2} 
        -2r\tilde{U} w^{T}d\theta\,dz
\nonumber\\
\label{SF:eq2}
	 &  & {}+ |RA(r\,d\theta + \beta\,dr + (\hat{\gamma}+ 
	 A^{-1}R^{T}\tilde{\Gamma})\,dz)|^{2}.
\end{eqnarray}

Now recall that $\Gamma= \Gamma(z,\rho,\zeta)$ is angular with respect to 
the $(\rho,\zeta)$ coordinates, and note that the pullback of $\zeta^{T}\Gamma=0$
simplifies using Eqn.~(\ref{zeta:eq}) to
\[
	0 = f^{-1}\,\theta^{T}A^{-1}R^{T}\tilde{\Gamma},
\]
where $\tilde{\Gamma}= \Phi^{*}(\Gamma) = 
\Gamma(z,\rho(z,r,\theta),\zeta(z,r,\theta))$.
This shows that the vector $A^{-1}R^{T}\tilde{\Gamma}$ is purely angular 
in the $(r,\theta)$ coordinates, and  the final term of Eqn.~(\ref{SF:eq2}) becomes
\[
	|r\,d\theta + \beta\,dr + (\hat{\gamma}
	+ A^{-1}R^{T}\tilde{\Gamma})\,dz|^{2}.
\]
Note also by Eqn.~(\ref{ATA:eq2}) that
\begin{equation}
	A^{-1}R^{T}\tilde{\Gamma} = \Theta A^{-1}R^{T}\tilde{\Gamma}
	= \Theta A^{T}A\Theta A^{-1}R^{T}\tilde{\Gamma}
	= \Theta A^{T}R^{T} \tilde{\Gamma}.
\label{Gtilde:eq}
\end{equation}
Introducing the angular vector 
\begin{equation}
	\gamma := \hat{\gamma}+ A^{-1}R^{T}\tilde{\Gamma}-\tilde{U}\Theta w
	= \hat{\gamma}+\Theta (A^{T}R^{T}\tilde{\Gamma}-\tilde{U}w)	
\label{gamma:def}
\end{equation}
and noting that $w^{T}A^{-1}=fw^{T}$,
the pullback metric becomes
\begin{eqnarray}
	\Phi^{*}(ds^{2}_{SF}) & = & {}|r\,d\theta+\beta\,dr + \gamma\,dz|^{2}
	-2 \tilde{U}f(1+r\theta^{T}t_{1})\,dr\,dz
\nonumber\\
	 &  & {}-2 \tilde{U}(\tilde{V}+ rf\theta^{T}t_{0} + {\textstyle 
	 \frac{1}{2}} \tilde{U}|\Theta w|^{2} - fw^{T}R^{T}\tilde{\Gamma})\,dz^{2}.
\label{Phi*ds2:ans}
\end{eqnarray}
Comparing this metric with the general NQS metric Eqn.~(\ref{ds2:nqs}), 
which may be written in 3-vector notation as 
\begin{equation}
	ds^{2}_{NQS} = -2u\,dz\,(dr+v\,dz) + |rd\theta+\beta\,dr+\gamma\,dz|^{2},
\label{NQSb:def}
\end{equation}
we obtain the main transformation result for shear-free metrics.

\begin{proposition}\label{genNQS:prop}
Suppose $\Phi\in C^{\infty}(\Omega\times S^{2},\bR\times\cCo)$ 
for some domain $\Omega\subset \bR^{2}$ and $\Phi$ 
is quasi-spherical with respect 
to the expanding shear-free NQS metric $ds^{2}_{SF}$ 
given by Eqn.~(\ref{ds2SF:def}), with null coordinate $z$. Then  
$\Phi$ is described by Eqn.~(\ref{Phi:def2}) with Lorentz boost and 
rotation functions $w\in C^{\infty}(\Omega,\bR^{3})$, $R\in 
C^{\infty}(\Omega,SO(3))$, and the pullback $\Phi^{*}(ds^{2}_{SF})$ is 
given by Eqn.~(\ref{Phi*ds2:ans}).
Defining the pullbacks $\tilde{U}=\Phi^{*}(U)$, $\tilde{V}=\Phi^{*}(V)$, 
$\tilde{\Gamma}=\Phi^{*}(\Gamma)$ and the
derived vectors $s_{0},s_{1},t_{0},t_{1}$
in terms of $w,R$ via 
Eqs.~(\ref{S1:def}),(\ref{s1:def}),(\ref{t1:def}), and 
(\ref{S0:def}),(\ref{s0:def}),(\ref{t0:def}), the NQS 
parameters $u,v,\beta,\gamma$ of the metric $\Phi^{*}(ds^{2}_{SF})$
are given explicitly by
\begin{eqnarray}
\label{u:eq2}
	u & = & \tilde{U}f(1+r\theta^{T}t_{1})
\\
\label{v:eq2}
	uv & = & \tilde{U}(\tilde{V}+ rf\theta^{T}t_{0} + {\textstyle 
	 \frac{1}{2}}\tilde{U} |\Theta w|^{2} - fw^{T}R^{T}\tilde{\Gamma})
\\
\label{beta:eq2}
	\beta & = &  r\Theta t_{1} + r \theta \times s_{1},
\\
\label{gamma:eq2}
	\gamma & = &r\Theta t_{0}+r\theta\times s_{0} 
	           +A^{-1}R^{T}\tilde{\Gamma}-\tilde{U}\Theta w.
\end{eqnarray}
Moreover, $\Phi$ is a diffeomorphism if the vector $t_{1}(z,r)$ defined by 
Eqn.~(\ref{t1:def}) satisfies 
\begin{equation}
	r\,|t_{1}| < 1,\quad \forall\ (z,r)\in \Omega.
\label{diffeo:eq}
\end{equation}
\end{proposition}

\proof  Because $ds^{2}_{SF}$ is non-degenerate (by assumption),
$\Phi$ will be a diffeomorphism iff the pullback is also non-degenerate,
and this holds exactly when $u$, the coefficient of $dz\,dr$ in 
$\Phi^{*}(ds^{2}_{SF})$, is non-zero. But $\tilde{U}\ne 0$ by assumption, 
and $f = \sqrt{1+|w|^{2}} + \theta^{T}w >0$ for all $w\in \bR^{3}$, 
$\theta\in S^{2}$, so the condition reduces to $1+r\theta^{T}t_{1} >0$.  
Clearly this holds for all $\theta\in S^{2}$ if and only if 
Eqn.~(\ref{diffeo:eq}) is satisfied.  All other 
statements of the proposition follow from previous computations.
\QED
 
\section{Examples}
\subsection{Spherically symmetric spacetimes}
The metric form 
\begin{equation}
	ds^{2}_{SS} = -2U\,dz(dr + V\,dr) + r^{2}|d\theta|^{2},
\label{ds2SS:def}
\end{equation}
with $U,V$ functions of $(z,r)$ only,
includes the Schwarzschild metric as the special case $U=1$, 
$V=\frac{1}{2}(1-2M/r)$, $M\in \bR$.  The geometric mass function 
$m=\frac{r}{2}(1-g^{ab}r_{,a}r_{,b})$ for the general 
metric (\ref{ds2SS:def}) is given by
\begin{equation}
	2m(z,r) = r(1-2V/U).
\label{SSmass:def}
\end{equation}
Again switching from $(r,\theta)$ to $(\rho,\zeta)$ coordinates in 
Eqn.~(\ref{ds2SS:def}), 
Proposition \ref{genNQS:prop} describes $ds^{2}_{SS}$ in  general Lorentz 
transformed NQS coordinates, with in particular $\tilde{\Gamma}=0$ and
\begin{equation}
	\tilde{U}(z,r,\theta) = U(z,\rho(z,r,\theta)),\qquad 
	\tilde{V}(z,r,\theta) = V(z,\rho(z,r,\theta)),
\label{SSuvt:eq}
\end{equation}
where $\rho(z,r,\theta) = rf(\theta;w(z,r)) = r(b+\theta^{T}w)$, 
$b=\sqrt{1+|w|^{2}}$.
In general the angular dependence of $u$ and $uv$ will be rather 
complicated, due to the effects of $\rho$-dependence of $\tilde{U},\tilde{V}$ in 
Eqn.~(\ref{SSuvt:eq}).

In the special case of the Schwarzschild metric $\tilde{U}=1$,
$2\tilde{V}=1-2M/\rho$, and the 
fields $\beta,\gamma$ given by Eqs.~(\ref{beta:eq2}), (\ref{gamma:eq2}) with
$\tilde{\Gamma}=0$ are both pure $\ell=1$ spin-1 spherical harmonics, and $u,v$ 
satisfy
\begin{eqnarray}
\label{schwu:eq}
	u & = & (b+\theta^{T}w)(1+r\theta^{T}t_{1}) = f(1+r\theta^{T}t_{1})
\\
\label{schwv:eq}
	2uv & = & {}-\frac{2M}{r(b+\theta^{T}w)} 
	+(b+\theta^{T}w)(b-\theta^{T}w+2r\theta^{T}t_{0}) .
\end{eqnarray}
If the boost $w\in\bR^{3}$ is constant and $R=I$, then we obtain the rather simple
NQS metric parameters
\begin{equation}
	\begin{array}{rlrl}
		\beta & =\ 0,\qquad& \gamma& = \ -\Theta w,
	\\
		u & =\ b+\theta^{T}w,\qquad& 2v 
		& =\ b-\theta^{T}w-\frac{2M}{r(b+\theta^{T}w)^{2}}\ ,
	\end{array}
\label{schwEg:eq}
\end{equation}
which describe the Schwarzschild spacetime in rigidly boosted coordinates.

Since in general the functions $w(z,r)\in\bR^{3}$, $R(z,r)\in SO(3)$ are arbitrary, 
subject only to smoothness and the size condition Eqn.~(\ref{diffeo:eq}), 
in order to construct challenging exact solutions for numerical relativity
benchmarking, we might choose $w,R$ in any reasonable manner, 
keeping $w=0$ and $R=I$ in 
regions where we wish the solution to remain explicitly equal to the 
standard Schwarzschild metric.  Note that asymptotic decay conditions may 
also be readily determined: for example the natural conditions 
\[
	w_{,r}=O(r^{-2})\quad w_{,z}=O(r^{-1})
\]
and similarly for $R$, give bounded $u,v$ and $\gamma$ with $\beta\to0$ as
$r\to\infty$.

\subsection{Accelerated Minkowski metric}
By moving the base point of the standard future light cone along a timelike 
curve in  Minkowski space, we may construct another class of shear-free 
NQS metrics.  The Minkowski metric associated with an accelerated null cone
foliation 
was discussed in \cite{NU63}, using a special choice of affine parameter 
on the null rays to determine the radius function.
Let $z\mapsto (p(z),\tau(z))\in\bR^{3,1}$ be a future-timelike curve in
Minkowski space and denote the tangent vector by $(\dot{p},\dot{\tau})$,
with $(\dot{-})$ indicating $d/dz$.  Two possible normalisations for $z$ are
$\dot{\tau} = 1$ and $\dot{\tau} = \sqrt{1+|\dot{p}|^{2}}$.
Define $\Psi:\bR^{3,1}\to\bR^{3,1}$ by 
\begin{equation}
	\Psi(z,r,\theta) = \left[
	\begin{array}{c}
		X  \\ T
	\end{array}
	\right] = \left[
	\begin{array}{c}
		  p(z)+r\theta \\ \tau(z)+r
	\end{array}\right]
\label{Psi:def}
\end{equation}
where $(z=t-r,r,\theta)$ are null-polar coordinates and $(X,T)$ are rectangular
coordinates on $\bR^{3,1}$.  Note that $\Psi$ maps the future null cone $z=const$ 
to the future null cone based at $(x,t)=(p(z),\tau(z))$.
The accelerated Minkowski metric 
$	ds^{2}_{AM} := \Psi^{*}(-dT^{2}+|dX|^{2})$ may be written
\begin{eqnarray}
	ds^{2}_{AM}  & = & 
	-(\dot{\tau}dz+dr)^{2}+ |\dot{p}dz + \theta dr + r d\theta|^{2}
\nonumber\\
	 & = & -2 (\dot{\tau}-\theta^{T}\dot{p})\,dz\,
	 (dr+{\textstyle \frac{1}{2}}(\dot{\tau}+\theta^{T}\dot{p})\,dz)
	 +|r d\theta + \Theta\dot{p}\,dz|^{2},
\label{ds2AM:eq}
\end{eqnarray}
which is a metric in the shear-free NQS form (\ref{ds2SF:def}), with 
coefficient functions
\begin{eqnarray}
\label{AMU:def}
	U & = & U(z,\theta) = \dot{\tau}(z)-\theta^{T}\dot{p}(z)
\\
\label{AMV:def}
	V & = & V(z,\theta) = {\textstyle \frac{1}{2}}(\dot{\tau}(z) + 
	\theta^{T}\dot{p}(z))
\\
\label{AMGam:def}
	\Gamma & = & \Gamma(z,\theta) = \Theta\dot{p}(z).
\end{eqnarray}
Note that the timelike condition $\dot{\tau}>|\dot{p}|$ ensures that $U,V$ 
are both strictly positive.

Proposition \ref{genNQS:prop} gives the NQS coefficients for the 
Lorentz-transformed metric $\Phi^{*}(ds_{AM}^{2})$ (with $ds^{2}_{AM}$ 
written in $(z,\rho,\zeta)$ as before),
and we may  simplify as follows.
The shear $\beta$ is given simply by Eqn.~(\ref{beta:eq2}) and because
$\Theta A^{T}R^{T}\tilde{\Gamma} = \Theta A^{T}R^{T}\dot{p}$, we find  that
\begin{equation}
	\gamma = r(\Theta t_{0} + \theta\times s_{0}) 
	+ \Theta(-\dot{\tau}w + WR^{T}\dot{p}).
\label{AMgamma:eq}
\end{equation}
Since $f\zeta = R(w+W\theta)=RA^{-1T}\theta$, we have
\[
	\tilde{U} = \dot{\tau} -\zeta^{T}\dot{p} = 
	\dot{\tau}-f^{-1}(w+W\theta)^{T}R^{T}\dot{p}
\]
and
\[
	\tilde{V}= {\textstyle \frac{1}{2}}(\dot{\tau}+\zeta^{T}\dot{p})
	=  {\textstyle \frac{1}{2}}(\dot{\tau}+f^{-1}(w+W\theta)^{T}R^{T}\dot{p})
\]
and $\tilde{\Gamma}=(I-\zeta\zeta^{T})\dot{p}$.
This gives immediately that
\begin{equation}
	u = (1+r\theta^{T}t_{1}) \,(f\dot{\tau}-(w+W\theta)^{T}R^{T}\dot{p}),
\label{AMu:eq}
\end{equation}
and after some computations, 
\begin{equation}
	2v = \frac{1}{1+r\theta^{T}t_{1}}
	    \left((b-\theta^{T}w)\dot{\tau}-(w-W\theta)^{T}R^{T}\dot{p} 
	    + 2 r\theta^{T}t_{0} \right).
\label{AMv:eq}
\end{equation}
Notice that the spherical harmonic decompositions of $u, uv$ contain terms 
with $\ell=0,1,2$ whereas $\beta,\gamma$ 
are both pure $\ell=1$.  

The coordinate system constructed in \cite{NU63} 
corresponds to an NQS metric constructed from the 
choices $R=I$, $w= \dot{p}(z)$, with proper time 
normalisation of $(p(z),\tau(z))$.
The metric Eqn.~(12) in \cite{NU63} may be transformed to NQS form with
NQS parameters
$\dot{\tau}=b$, $u=1$, $\beta=0$, $2v=1+2r\theta^{T}t_{0}$, 
$\gamma=r\Theta t_{0} + r \theta\times s_{0}$, where
$t_{0}= W^{-1}\ddot{p}$, $s_{0}= (b+1)^{-1}\ddot{p}\times\dot{p}$.
Thus if the acceleration $\ddot{p}$ is non-zero then  both $v,\gamma$ will
be unbounded as $r\to\infty$.
The transformation between the NU and NQS coordinates amounts to 
redefining the angular variables (the null cones and quasi-spheres are 
unchanged), and has the effect of moving the conformal isometry $P$ (cf 
Eqn.~(13) of \cite{NU63}) to the NQS field $\gamma$. 

Alternatively, the choice $w=R^{T}\dot{p}$, $\dot{\tau}=b$,
 with $\dot{R}=\frac{1}{b+1}(\dot{p}\ddot{p}^{T}-\ddot{p}\dot{p}^{T})R$ 
gives a Fermi-Walker transported spatial frame.  In this case we have 
$*S_{0}=\frac{b}{b+1}R^{T}\dot{p}\times\ddot{p}$, $s_{0}=0$, $t_{0}= 
W^{-1}R^{T}\ddot{p}$, and $u=1$, $2v=1+2r\theta^{T}t_{0}$, $\beta=0$, 
$\gamma=r\Theta t_{0}$.  I am indebted to Andrew Norton for this computation.
Note that in this case, the parameters $(\tau,p)$ and $(w,R)$ may be 
recovered from the metric data $t_{0}(z)$ by solving $\dot{w}=bt_{0}$, 
$\dot{R}=(b+1)^{-1}R(wt_{0}^{T}-t_{0}w^{T})$ and $\dot{p}=Rw$, with initial 
conditions $w(z_{0})=0$, $R(z_{0})=I$ and $p(z_{0})=0$ corresponding to an 
initial frame at rest.

\subsection{Robinson-Trautman metrics}
It was shown by Robinson and Trautman \cite{RT62} that vacuum spacetimes
which contain a null geodesic congruence which is  
hypersurface-orthogonal, expanding
and shear-free have particularly simple structure.  A coordinate 
transformation \cite{DIW69} brings such metrics to the  NQS form
\begin{equation}
	ds^{2}_{RT} = -2U\,dz\,(dr + {\textstyle \frac{1}{2}} 
	(\Delta_{0}U + U - 2MU^{-2}/r)\,dz)
	+ |rd\theta - U_{,\theta}^{T}dz|^{2}
\label{ds2RT:def}
\end{equation}
where $M\in \bR$ is constant, $\Delta_{0}=\Delta_{S^{2}}$ is the standard 
metric Laplacian on $S^{2}$ and $U=U(z,\theta)$ is independent of $r$.
The vacuum Einstein equations are satisfied by $ds^{2}_{RT}$ if $U$ satisfies the 
nonlinear parabolic equation
\begin{equation}
	12M\,\frac{\partial U}{\partial z} + U^{3}\Delta_{0} K = 0
\label{RTeqn}
\end{equation}
where $M\ne0$ and $K=U^{2}(\Delta_{0}\log U + 1)$ is the Gauss curvature 
of the metric $U^{-2}ds^{2}_{0}$ conformal to the standard metric 
$ds^{2}_{0}$ on $S^{2}$.
If $M=0$ then the metric reduces to the accelerated Minkowski metrics 
considered above \cite{RT62}.  
A global existence theorem for Eqn.~(\ref{RTeqn}) has been given by 
Chru\'sciel \cite{Chrusciel91}.

The RT metric has NQS parameters
$U$, 
$V={\textstyle \frac{1}{2}} (\Delta_{0}U + U + 2MU^{-2}/r)$, 
$\beta=0$ and $\Gamma=-U_{,\theta}^{T}$, and after changing 
Eqn.~(\ref{RTeqn}) over to $(z,\rho,\zeta)$ coordinates, Proposition 
\ref{genNQS:prop} describes the effect of a general Lorentz deformation of 
the coordinates.  In particular, since $\tilde{\Gamma}(z,r,\theta) = 
-U_{,\zeta}^{T}(z,\zeta(z,r,\theta))$, using the identity
\[
	\tilde{U}_{,\theta} = D_{\theta}U(z,\zeta) 
	= f^{-1}U_{,\zeta}RA\Theta
\]
and Eqn.~(\ref{Gtilde:eq}), we may simplify terms involving $\tilde{\Gamma}$:
\begin{eqnarray*}
	\Theta A^{T}R^{T}\tilde{\Gamma} & = & -f \tilde{U}_{,\theta}^{T},
\\
	fw^{T}R^{T}\tilde{\Gamma} & = & w^{T}A^{-1}R^{T}\tilde{\Gamma}= 
	-f w^{T} \tilde{U}_{,\theta}^{T}.
\end{eqnarray*}
Thus in the RT case Eqs.~(\ref{v:eq2}),(\ref{gamma:eq2}) become
\begin{eqnarray}
\label{RTuv:eq}
	uv & = & \tilde{U}(\tilde{V} + rf\theta^{T}t_{0} + {\textstyle \frac{1}{2}} 
	\tilde{U}|\Theta w|^{2} - f w^{T}\tilde{U}_{,\theta}^{T}),
\\
\label{RTgamma:eq}
	\gamma & = & r\Theta t_{0}+ r\theta\times s_{0} - (f\tilde{U})_{,\theta}^{T},
\end{eqnarray}
and $u=f\tilde{U}(1+r\theta^{T}t_{1})$, $\beta=r\Theta t_{1}+ r\theta\times s_{1}$.

Note that if the Lorentz deformation preserves $\beta=0$ then 
$w=w(z)$, $R=R(z)$ and $u=f\tilde{U}$,
and $\gamma$ will remain independent of $r$ only if $s_{0}=t_{0}=0$. 
This requires that $w,R$ are constant, and the transformed 
metric will again be in the explicit RT form of Eqn.~(\ref{ds2RT:def}).
This global Lorentz transformation may be used to normalise to zero the 
$\ell=1$ spherical harmonic components of $\lim_{z\to\infty}u(z,\theta)$
(or equivalently, of $\lim_{z\to\infty}\gamma(z,\theta)$) --- this 
transformation may be interpreted as defining an asymptotic rest frame for 
the RT spacetime \cite{Singleton90}, \cite{Chrusciel91}.

\section{Discussion}
In the case of vanishing shear $\sigma_{NP}=0$
(and assuming non-zero expansion $\rho_{NP}\ne0$ and spherical sections), 
we have seen that the null hypersurfaces are isometric to the standard 
cone $\cCo$ (Proposition \ref{Prop8}), and the residual freedom in the NQS 
gauge consists precisely of a Lorentz transformation at each quasi-sphere.
The transformed metric has NQS shear $\beta$ consisting purely of $\ell=1$
spherical harmonics, and conversely, if $\beta $ is pure $\ell=1$ then 
there is an inverse quasi-spherical map which transforms the metric into 
NQS form with $\beta=0$.  Thus the $\ell=1$ spherical harmonic components 
of the NQS shear $\beta$ are pure gauge.

Generalised Lorentz transformations preserving the condition $\beta=0$  have 
parameters $(w,R)$ depending only on $z$, since Eqn.~(\ref{beta:eq2}) 
combined with Eqs.~(\ref{wr:ode}),(\ref{Rr:ode}) show that $w_{,r}$ and 
$R_{,r}$ must vanish.
This remaining gauge freedom may be used to set the $\ell=1 $ components of 
$\gamma$ to zero at one fixed radius $r_{0}$ as follows.  The six $\ell=1$ 
coefficients of the terms $\Theta (A^{T}R^{T}\tilde{\Gamma}-\tilde{U}w)$ 
of $\gamma$ (cf.~Eqn.~(\ref{gamma:eq2})) form a nonlinear functional of 
$w,R$, so Lemma \ref{lemma2} may be used to solve for $w_{,z}, R_{,z}$ from 
$s_{0},t_{0}$, giving a system of ordinary differential equations
\[
	\frac{d}{dz} (w(z,r_{0}),R(z,r_{0})) = F(z,r_{0};w,R),
\]
where $F(z,r_{0};w,R)$ is linearly bounded in $w$.  Consequently there 
exists a solution which is global in $z$, which in turn ensures
(after applying the resulting Lorentz transformation) 
that $\gamma(z,r_{0})$ has vanishing $\ell=1$ components at each $z$.

Alternatively, it might be possible to use the gauge freedom $w(z),R(z)$ 
to normalise 
the $\ell=1$ components of $u(z,r_{0})$ using Eqn.~(\ref{u:eq2}), since 
$f=b+\theta^{T}w$ is pure $\ell=0,1$ and $t_{1}=0$ by the condition 
$\beta=0$.  Note that this remaining freedom is similar to that available 
in the Bondi and Newman-Unti gauges.
In any case, it is a plausible conjecture that the gauge freedom remaining
in the general NQS metric (\ref{ds2:nqs}) may be used to eliminate the
$\ell=1$ components of $\beta$, and that the freedom remains to make
a rigid Lorentz transformation on each null hypersurface.

The interpretation of $\ell=1$ spherical harmonic components as gauge terms 
has also been noted in the construction of the Regge-Wheeler-Zerilli 
equations for linearised perturbations of Schwarzschild 
\cite{RW57},\cite{Zerilli70},\cite{Moncrief74}.  The gauge-invariant 
quantities satisfying the RWZ equations are constructed from 
$\ell\ge2$ components of the metric 
perturbations.  Furthermore,  
one quantity constructed from the $\ell=1$ 
components represents (non-dynamic) angular momentum 
\cite{Moncrief74},\cite{Spillane94} arising from the Kerr perturbation of the 
Schwarzschild metric --- this quantity
corresponds to the linearised limit of the odd (rotational) $\ell=1$ 
component of $\partial_{z}(\beta/r)-\partial_{r}(\gamma/r)$, 
and vanishes for the pure gauge variations constructed above.

\acknowledgements
{The support of the Institute for Mathematical Sciences at 
the Chinese University of Hong Kong during the preparation of this paper 
is gratefully acknowledged.  I also wish to thank Andrew Norton for his 
careful review of this manuscript, which has helped to clarify many of the 
computations.}

\appendix
\section{Shear-free spacetimes}
Let $\cN$ be a null hypersurface in some spacetime, with induced 
(degenerate) metric $ds^{2}_{\cN}=g_{\cN}$.  An {\em adapted null  frame} on 
$\cN$ is a pair of vector fields $(l,m)$ where $l$ is a degeneracy 
vector for $ds^{2}_{\cN}$, $m\in T\cN\otimes\bC$, and 
$(l,m+\bar{m},\imu(m-\bar{m}))$ form a real basis for $T\cN$, such that
\[
	g_{\cN}(m,m) = g_{\cN}(l,l) = g_{\cN}(l,m)=0,\quad g_{\cN}(m,\bar{m})=1.
\]
Using $\nabla$ to denote the ambient spacetime covariant derivative, we 
define the shear and expansion of $ds^{2}_{\cN}$ (with respect to the null 
adapted frame $(l,m)$) by
\begin{eqnarray}
\label{shear:def}
	\mbox{shear}\ = \sigma_{NP}& = & -g(\nabla_{m}m,l),
\\
\label{expansion:def}
	\mbox{expansion}\ = \rho_{NP} & = & -g(\nabla_{m}\bar{m},l).
\end{eqnarray}
Although we use the Newman-Penrose notation, the importance of the shear 
and expansion of a null geodesic congruence was known prior to
\cite{NP62} --- see \cite{Sachs61} for example.

\begin{lemma}
   Let $(l,m)$ be an adapted null frame for $\cN$, then the shear and 
   expansion depend only on $(l,m)$ and $ds^{2}_{\cN}$.  In particular we 
   have
   \begin{eqnarray}
   \label{shear2:eq}
   	\sigma_{NP} & = & -g_{\cN}(m,[l,m])
   \\
   \label{expansion2:eq}
   	\rho_{NP} & = & 
   	     -{\textstyle \frac{1}{2}}(g_{\cN}(m,[l,\bar{m}]) +
   	              g_{\cN}(\bar{m},[l,m])),
   \end{eqnarray}
   where $[l,m]$ is the Lie bracket.
\end{lemma}
\proof The identities (\ref{shear2:eq}),(\ref{expansion2:eq}) are easily 
verified since $[l,m]$, being the Lie bracket of vector fields tangent to 
the hypersurface $\cN$, is again tangent to $\cN$.
\QED

\begin{lemma}
   Suppose $(l',m')$ is a null adapted frame which presents the same 
   orientation of $\cN$ and the null generators as $(l,m)$.  There are real 
   functions $\alpha,\lambda,\mu\in C^{\infty}(\cN)$ such that
  \begin{equation}
	  m' = e^{\imu\lambda}m + \alpha l,\quad l'=e^{\mu}l,
  \label{framechange:eq}
  \end{equation}
   and the shear and expansion satisfy
   \begin{eqnarray}
   \label{shear3:eq}
   	\sigma_{NP}' & = & e^{\mu+2\imu\lambda}\sigma_{NP}
   \\
   \label{expansion3:eq}
   	\rho_{NP}' & = & e^{\mu}\rho_{NP}.
   \end{eqnarray}
   Consequently the conditions ``$\rho_{NP}\ne 0$, $\sigma_{NP}=0$ everywhere 
   on $\cN$'' are independent of the choice of adapted null frame, and we 
   may consider $\sigma_{NP}/\rho_{NP}$ as a section of a spin-2 complex 
   line bundle over $\cN$.
\end{lemma}
\proof
The representation (\ref{framechange:eq}) for the frame change follows 
directly from the orientation and orthogonality conditions.  The formula
\[	
   [m',l'] = e^{\mu+\imu\lambda}([m,l] +D_{m}\mu\,l - \imu D_{l}\lambda\,m)
	   + e^{\mu}(\alpha D_{l}\mu - D_{l}\alpha)\,l
\]
where $D_{l},D_{m} $ are the directional derivative operators,
leads directly to Eqs.~(\ref{shear3:eq}),(\ref{expansion3:eq}).
\QED

Using the foliation of $\cN$ by null generating curves, we may introduce adapted 
coordinates $(\rho,x^{3},x^{4})$  by requiring $(x^{3},x^{4})$ to be 
constant along the null generators, and then allowing $\rho$ to be any 
parameterisation of the null generators.  In such coordinates the metric 
becomes 
\[
	ds^{2}_{\cN} = h_{ab}dx^{a}dx^{b},
\]
where the indices $a,b$ have range $3,4$ and $h_{ab}=h_{ab}(\rho,x^{3},x^{4})$.
A natural choice of null frame is $l=\partial_{\rho}$ and 
$m=m^{a}\partial_{a}$, where $\partial_{\rho}, \partial_{3}, \partial_{4}$ 
are the coordinate tangent vectors.  
Introducing the cotangent vector $m_{a}dx^{a}$, where the $m_{a}, a=3,4$ are 
defined by the requirements $m^{a}m_{a}=0$, $\bar{m}^{a}m_{a}=1$, we have
\[
	h_{ab} = m_{a}\bar{m}_{b} + \bar{m}_{a}m_{b}.
\]
Direct computation using $[l,m]=\partial_{\rho}(m^{a})\partial_{a}$ gives 
the following expressions for the shear and expansion with respect to the 
coordinate-based null framing $(l,m)$:
\begin{eqnarray}
\nonumber
	\sigma_{NP} & = & m^{a}\partial_{\rho}(m_{a})
\\
\label{shear4:eq}
	 & = & {\textstyle\frac{1}{2}}m^{a}m^{b}\partial_{\rho}h_{ab}
\\
\nonumber
	\rho_{NP} & = & {\textstyle\frac{1}{2}}(\bar{m}^{a}\partial_{\rho}(m_{a})
	  + m^{a}\partial_{\rho}(\bar{m}_{a}))
\\
\label{expansion4:eq}
	 & = & {\textstyle\frac{1}{4}}h^{ab}\partial_{\rho}h_{ab},
\end{eqnarray}
where $[h^{ab}] = [h_{ab}]^{-1} = m^{a}\bar{m}^{b}+ \bar{m}^{a}m^{b}$.

If $\cN$ is shear-free and expanding then the metric on $\cN$ may be 
brought into explicitly NQS form.  It should be possible to extend this 
result to allow some non-zero shear, but the proof will be considerably more 
difficult.
\begin{proposition}\label{Prop8}
   Suppose $\cN$ is a null 3-manifold with everywhere vanishing shear and 
   non-zero expansion, and having spatial cross-sections which are topological 
   spheres.  Then there exist polar coordinates $(r,\vartheta,\varphi)$ on 
   $\cN\simeq \bR\times S^{2}$ 
   such that
   $ds^{2}_{\cN}=r^{2}(d\vartheta^{2}+ \sin^{2}\vartheta\,d\varphi^{2})$.
\end{proposition}
Note that the following argument may be easily adapted in case the spatial sections are 
not spheres.

\proof
Let $l=\partial_{\rho}$, $m=m^{a}\partial_{a}$ be a coordinate-based null 
frame for $\cN$.  By Eqn.~(\ref{shear4:eq}) and the shear-free condition we 
have
\[
	m^{a}m^{b}\partial_{\rho}h_{ab} =0.
\]
Now $\partial_{\rho}h_{ab}$ may be decomposed 
\[
	\partial_{\rho}h_{ab} = A (m_{a}\bar{m}_{b}+\bar{m}_{a}m_{b}) + 
	   B m_{a}m_{b} + \bar{B}\bar{m}_{a}\bar{m}_{b},
\]
where $A$ is real-valued and $B$ is complex-valued.  The shear-free 
condition shows that $B=0$ and the resulting equation 
$\partial_{\rho}h_{ab}=Ah_{ab}$ may be integrated along each null generator 
to give
\[
	h_{ab}(\rho,x) = \exp\left(\int^{\rho}_{\rho_{0}} A(s,x)ds\right) 
	h^{0}_{ab}(x),
\]
where $h^{0}_{ab}(x) = h_{ab}(\rho_{0}(x),x)$ is a fixed metric on $S^{2}$.
Now the Riemann Uniformisation Theorem \cite{CK93} shows there is a 
diffeomorphism $\Phi:S^{2}\to S^{2}$, $x\mapsto (\vartheta(x),\varphi(x))$, 
and a function $\phi\in C^{\infty}(S^{2},\bR^{+})$ such that
$h^{0}_{ab}dx^{a}dx^{b} 
= \phi^{2}(x)\Phi^{*}(d\vartheta^{2}+ \sin^{2}\vartheta\,d\varphi^{2})$.
Using the coordinates $(\vartheta,\varphi)$ to label the null generators 
gives the representation
\[
	ds^{2}_{\cN}= \exp\left(\int^{\rho}_{\rho_{0}} A\right) \phi^{2}
	(d\vartheta^{2}+ \sin^{2}\vartheta\,d\varphi^{2}).
\]
Now define the positive function $r=r(\rho,\vartheta,\varphi)$  by 
$r=\exp\left(\frac{1}{2}\int^{\rho}_{\rho_{0}} A\right) \phi$.
Since $4\rho_{NP}=h^{ab}\partial_{\rho}h_{ab} = 2A$, it follows that
\[
	\frac{\partial r}{\partial \rho} = r \rho_{NP} > 0,
\]
hence $r$ is a valid coordinate, and $ds^{2}_{\cN}$ takes the required form 
in the coordinates $(r,\vartheta,\varphi)$.
\QED


\end{document}